\documentclass[%
 reprint,
groupedaddress,
amsmath,amssymb,
aps,
pra,
]{revtex4-1}
\usepackage{graphicx}
\usepackage{dcolumn}
\usepackage{bm}
\usepackage{multirow}
\usepackage{float}
\usepackage[dvipdfm,colorlinks,linkcolor=blue, urlcolor=blue,anchorcolor=blue, citecolor=blue]{hyperref}
\begin{document}

\title{Three-level Dicke quantum battery}

\author{Dong-Lin Yang}
\affiliation{College of Physics and Electronic Engineering, Northwest Normal University, Lanzhou, 730070, China}
\author{Fang-Mei Yang}
\affiliation{College of Physics and Electronic Engineering, Northwest Normal University, Lanzhou, 730070, China}
\author{Fu-Quan Dou}
\email{doufq@nwnu.edu.cn}
\affiliation{College of Physics and Electronic Engineering, Northwest Normal University, Lanzhou, 730070, China}


\begin{abstract}
Quantum battery (QB) is the energy storage and extraction device that is governed by the principles of quantum mechanics. Here we propose a three-level Dicke QB and investigate its charging process by considering three quantum optical states: a Fock state, a coherent state, and a squeezed state. The performance of the QB in a coherent state is substantially improved compared to a Fock and squeezed states. We find that the locked energy is positively related to the entanglement between the charger and the battery, and diminishing the entanglement leads to the enhancement of the ergotropy. We demonstrate the QB system is asymptotically free as $N \rightarrow \infty$. The stored energy becomes fully extractable when $N=10$, and the charging power follows the consistent behavior as the stored energy, independent of the initial state of the charger.
\end{abstract}

\maketitle


\section{INTRODUCTION}
One of the most promising applications relevant to future quantum technologies is ``quantum batteries (QBs)", i.e., quantum mechanical systems for temporarily storing and then extracting energy \cite{PhysRevE.87.042123}. The QB can exploit the quantum resources (such as quantum entanglement or quantum coherence) to obtain more efficient work extraction and faster charging processes with respect to classical schemes \cite{campaioli2018quantum}. Considerable attention has been mostly focused on the development of QBs with superior performance \cite{Binder_2015,PhysRevLett.120.117702,PhysRevLett.122.047702,
PhysRevB.98.205423,PhysRevB.99.205437,PhysRevA.97.022106,PhysRevB.100.115142,PhysRevA.100.043833,
PhysRevLett.118.150601,Friis2018precisionwork,PhysRevB.99.035421,PhysRevE.99.052106,Alicki2019,PhysRevE.100.032107}. It is pivotal to continue research on the construction of QB models \cite{Binder_2015,PhysRevLett.120.117702,PhysRevA.97.022106,PhysRevLett.122.047702,PhysRevB.98.205423,
PhysRevB.99.205437,PhysRevB.100.115142,PhysRevA.100.043833,PhysRevE.100.032107,PhysRevLett.124.130601,PhysRevLett.125.236402,
ito2020collectively,PhysRevB.102.245407,Chen2020,Rosa2020,Moraes2021,PhysRevA.103.052220,Dou2020,Dou2021,
PhysRevE.101.062114,PhysRevA.104.032606,PhysRevA.106.022618,PhysRevA.101.032115,PhysRevA.103.033715,
PhysRevResearch.4.013172,dou2022charging,PhysRevE.104.024129,PhysRevE.105.054115,PhysRevA.107.032203,PhysRevA.107.042419,PhysRevB.105.115405,PhysRevA.107.023725,guo2023analytically},
the roles of quantum resources
\cite{PhysRevE.87.042123,PhysRevLett.118.150601,PhysRevLett.122.047702,PhysRevE.94.052122,PhysRevA.107.022215,
PhysRevA.106.062609,PhysRevA.106.032212,PhysRevLett.111.240401,PhysRevE.102.052109,Gumberidze2019,PhysRevA.104.L030402,
PhysRevB.104.245418,PhysRevLett.128.140501,centrone2021charging,PhysRevE.106.054107,PhysRevResearch.2.023113},
and the effects of initial state
\cite{Crescente_2020,PhysRevLett.122.047702,PhysRevA.106.062609,delmonte2021characterization,PhysRevA.104.043706,
PhysRevA.105.022628,PhysRevA.105.023718,LipkaBartosik2021secondlawof,PhysRevResearch.2.023095,PhysRevE.105.064119} to optimize the performance of QBs. In addition, experimental results have also shown advances towards the exploration of QBs \cite{quach2020,Hu2022,PhysRevA.106.042601,wenniger2022coherence,gemme2022ibm,Zheng2022,PhysRevA.107.L030201}.

In the quest for high-performance of QBs, several effort has been dedicated to the impact of quantum entanglement in extractable work \cite{Binder_2015,PhysRevLett.122.047702,PhysRevE.94.052122,PhysRevA.107.022215,PhysRevA.106.062609} and charging power \cite{Binder_2015,PhysRevLett.120.117702,PhysRevLett.111.240401,PhysRevLett.128.140501}. Alicki and Fannes proposed that entangling unitary controls perform better in work extraction capabilities from a QB, when compared to local controls \cite{PhysRevE.87.042123}. Entanglement generation can lead to a speed-up in the process of work extraction, which was demonstrated in Ref.~\cite{PhysRevLett.111.240401}. Afterward, the collective charging scheme of the ensemble of quantum cells induce a quantum advantage in the charging power \cite{Binder_2015,PhysRevLett.118.150601,PhysRevLett.120.117702}. Another effort is to investigate the different initial states how to influence the performance of QBs \cite{PhysRevLett.122.047702, PhysRevA.106.062609,
delmonte2021characterization,PhysRevE.105.064119}. A Tavis-Cummings (TC) QB considered three different initial states of the charger and confirmed that the coherent state is optimal for energy extraction in Ref.~\cite{PhysRevLett.122.047702}. Despite such progress, it remains vital to further explore the correlation between entanglement, initial states, and extractable work.

A three-level system interacting with light as it is related to an important class of quantum optical phenomena, including electromagnetically induced transparency \cite{PhysRevLett.66.2593,RevModPhys.77.633}, lasing without inversion \cite{PhysRevLett.62.2813,J_Mompart2000-vb}, as well as methods such as stimulated Raman adiabatic passage (STIRAP) \cite{10.1063/1.458514,RevModPhys.89.015006}. It is desirable to
take the three-level system as quantum cells of QBs. A recent work proposed the three-level QB, using the STIRAP technique to bypass the undesired spontaneous discharging and facilitate efficient energy transfer between the ground state and the maximum excited state \cite{PhysRevE.100.032107}. Later on, a closed-loop three-level QB was introduced where closed-contour interaction can effectively improve the charging performance \cite{Dou2020}. Furthermore, a three-level QB utilizing the shortcut to adiabaticity realized the highly efficient charging and discharging processes \cite{Dou2021}. Experimentally, three-level QBs based on superconducting circuits were also reported, including the transmon qutrit QB \cite{Hu2022} and the Xmon qutrit QB \cite{Zheng2022}.

Another paradigmatic model, in which a collection of the two-level systems coupled to a quantized single-mode light field, called the Dicke model \cite{PhysRev.93.99}, has extensive applications in QBs \cite{PhysRevLett.120.117702,zhang2018enhanced,PhysRevB.102.245407,PhysRevB.105.115405,quach2020}. The extension of the two-level Dicke model to the three-level system has been studied \cite{Sung1979,Crubellier1985,Crubellier1986,PhysRevA.79.053622,PhysRevLett.121.173602,PhysRevA.84.053856,
PhysRevA.86.063822,PhysRevA.87.023813,PhysRevA.87.023805,PhysRevA.107.033711,PhysRevA.104.063705,
PhysRevLett.127.253601,PhysRevResearch.4.013100,PhysRevA.105.053702,PhysRevLett.128.153601},
such as the subradiance \cite{Sung1979,Crubellier1985,Crubellier1986,PhysRevA.79.053622,PhysRevLett.121.173602}, superradiant phase transitions \cite{PhysRevA.84.053856,PhysRevA.86.063822,PhysRevA.87.023813,PhysRevA.87.023805,PhysRevA.107.033711}, time crystalline order \cite{PhysRevA.104.063705,PhysRevLett.127.253601}, and enantiodetection of chiral molecules \cite{PhysRevResearch.4.013100}. Encouraged by the new quantum optical phenomena beyond the two-level case in the context of three-level atomic structure, it is natural to inquire whether the implementation of the three-level Dicke model can lead to enhanced performance of QBs.

In this work, we propose a three-level Dicke QB composed by $N$ $\Xi$-type three-level atoms and a single-mode cavity. The QB's charging process is discussed with three
quantum optical states, i.e, a Fock state, a coherent state, and a squeezed state. We introduce the von Neumann entropy to characterize the entanglement between the charger and the battery and investigate the correlation between the locked energy and the entanglement. We are concerned with the dependence of the QB’s maximum stored energy and the maximum charging power on the cavity-atom coupling and introduce the quantum phase transitions, Wigner function, and photon distribution to analyze the behavior of the maximum stored energy. In addition, we also analyze the effect of $N$ on the maximum stored energy,
\begin{figure}[htbp]
\centering
\includegraphics[width=0.45\textwidth]{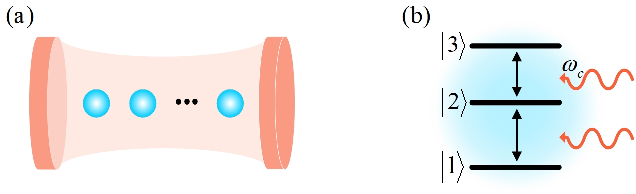}
\caption{\label{fig.1}(a) Sketch of a three-level Dicke QB composed by $N$ identical and independent three-level atoms coupled a single cavity mode. (b) A $\Xi$-type three-level atom with three energy levels labeled as $|1\rangle$, $|2\rangle$, and $|3\rangle$, respectively. The transitions $|1\rangle \leftrightarrow |2\rangle$ and $|2\rangle \leftrightarrow |3\rangle$ are mediated by a single cavity mode with frequency $\omega_{c}$.}
\end{figure}
maximum charging power, entanglements and ratios. Finally, we show the asymptotic freedom
of the QB in the $N \rightarrow \infty$ limit.

This work is organized as follows. Sec \ref{section2} introduces the three-level Dicke QB and defines the relevant physical quantities to characterize the behavior of the QB. Then we discuss the charging process of the QB for three different initial states and analyze the influence of the entanglement on the QB in Sec. \ref{section3}. Finally, we give a summary in Sec. \ref{section4}.

\section{MODEL} \label{section2}
We consider a QB modeled as $N$ identical and independent $\Xi$-type three-level atoms coupled to a single-mode cavity, as shown in Fig.~\ref{fig.1}. The QB system can
be described by the following Hamiltonian
\cite{Sung1979,Crubellier1985,Crubellier1986,PhysRevA.84.053856}
($\hbar = 1$ throughout this work):
\begin{eqnarray}\label{H(t)}
H(t)=H_{A}+H_{B}+\lambda(t)H_{I},
\end{eqnarray}
where the time-dependent parameter $\lambda(t)$ describes the charging time interval, which we assume to be given by a step function equal to $1$ for $t\in[0,T]$ and zero elsewhere. $H_{A,B,I}$ are the Hamiltonians of the charger, battery and interaction terms, respectively, and with the following forms:

\begin{eqnarray}\label{H(A)}
H_{A}=\omega_{c}\hat{a}^{\dagger} \hat{a},
\end{eqnarray}
\begin{eqnarray}\label{H(B)}
H_{B}=\sum_{i=1}^{3}\omega_{i}\hat{A}_{ii},
\end{eqnarray}
\begin{eqnarray}\label{H(I)}
\begin{split}
H_{I}=&\frac{g_{12}}{\sqrt{N}}(\hat{a}^{\dagger} + \hat{a})(\hat{A}_{12}+\hat{A}_{21})\\
&+\frac{g_{23}}{\sqrt{N}}(\hat{a}^{\dagger} + \hat{a})(\hat{A}_{23}+\hat{A}_{32}).\\
\end{split}
\end{eqnarray}
Here $\hat{a}$ ($\hat{a}^{\dagger}$) annihilates (creates) a photon in the cavity with frequency $\omega_{c}$. $\hat{A}_{ij}=\sum_{k=1}^{N}|i_{k}\rangle\langle j_{k}|$$(i,j=1,2,3)$ are the collective operators with $|i_{k}\rangle$ denoting the $i$th level of the $k$th atom. The energies of the three states are $\omega_{1} < \omega_{2} < \omega_{3}$. $g_{ij}$ is the cavity-atom coupling strength between states $|i\rangle$ and $|j\rangle$.

We consider the charging process of the three-level Dicke QB in a closed quantum system. The $N$ $\Xi$-type three-level atoms are prepared in ground state $|1\rangle$ and coupled to a single-mode cavity in the three typical quantum optical states : a Fock state, a coherent state, and a squeezed state, all having the same input energy $2N\omega_{c}$. Thus, the initial state of the total system is
\begin{equation}\label{psi0}
|\varPsi(0)\rangle=|\varPsi(A)\rangle\otimes|\underbrace{1,1,...,1}_{N}\rangle,
\end{equation}
$|\varPsi(A)\rangle$ is the cavity state with full energy.

In our charging protocol, QB will start charging when the classical parameter $\lambda(t)$ is nonzero. The wave function of the system evolves with time, i.e.,
\begin{eqnarray}\label{psit}
|\varPsi(t)\rangle=U|\varPsi(0)\rangle=e^{-iHt/\hbar}|\varPsi(0)\rangle.
\end{eqnarray}

The stored energy of QB at time $t$ is given by
\begin{eqnarray}\label{E(t)}
E_{B}(t)=\mathrm{Tr}[H_{B}\rho_{B}(t)],
\end{eqnarray}
where $\rho_{B}(t) = \mathrm{Tr_{A}}[\rho_{AB}(t)]$ is the reduced density matrix of the battery. However, $E_{B}(t)$ cannot be wholly extracted from the battery, which is known as the second law of thermodynamics. Therefore, a proper measure of the extractable work is provided by the ergotropy \cite{Allahverdyan2004}
\begin{eqnarray}\label{e(t)}
\mathcal{E}_{B}(t)=E_{B}(t)-\min_{U}\mathrm{Tr}[H_{B}U\rho_{B}(t)U^{\dagger}].
\end{eqnarray}
Consider the spectral decompositions of the $\rho_{B}$ and $H_{B}$ as
$\rho_{B}=\sum_{n}r_{n}|r_{n}\rangle\langle{r_{n}|}$ and $H_{B}=\sum_{n}\epsilon_{n}|\epsilon_{n}\rangle\langle{\epsilon_{n}|}$ so that $r_{0} \geqslant r_{1} \geqslant \cdots $ and $\epsilon_{0} \leqslant \epsilon_{1} \leqslant \cdots $. The passive counterpart of $\rho_{B}$ is $\widetilde{\rho}=\sum_{n}r_{n}|\epsilon_{n}\rangle\langle{\epsilon_{n}|}$, and its mean energy is unextractable (locked energy) and given by
\begin{eqnarray}\label{Erho}
E_{\widetilde{\rho}}(t)=\mathrm{Tr}[H_{B}\widetilde{\rho}(t)]=\sum_{n}r_{n}\epsilon_{n}.
\end{eqnarray}
It corresponds to the second term on the right-hand side of Eq. (\ref{e(t)}), i.e.,
$\min_{U}\mathrm{Tr}[H_{B}U\rho_{B}(t)U^{\dagger}]=\sum_{n}r_{n}\epsilon_{n}$.

The average charging powers of $E_{B}(t)$ and $\mathcal{E}_{B}(t)$ are given by
\begin{eqnarray}\label{P(t)}
P_{B}(t)=E_{B}(t)/t,
\end{eqnarray}
\begin{eqnarray}\label{P1(t)}
\mathcal{P}_{B}(t)=\mathcal{E}_{B}(t)/t.
\end{eqnarray}

We are interested in exploring the correlation between the entanglement and the ergotropy. Since in our model the quantum state of total system (charger plus battery) remains pure at all times and the battery state $\rho_{B}(t)$ will be mixed because of its entanglement with the charger during the time evolution, the entanglement between the charger and battery can be characterized by the $von$ $Neumann$ $entropy$ \cite{PhysRevLett.80.2245,PhysRevA.53.2046,RevModPhys.80.517} of the battery's reduced density matrix $\rho_{B}(t)$. The von Neumann entropy is defined by
\begin{eqnarray}\label{S(t)}
S(t) = -\mathrm{Tr}[\rho_{B}(t)\log_{2}\rho_{B}(t)].
\end{eqnarray}
\begin{figure*}[htbp]
\centering
\includegraphics[width=0.8\textwidth]{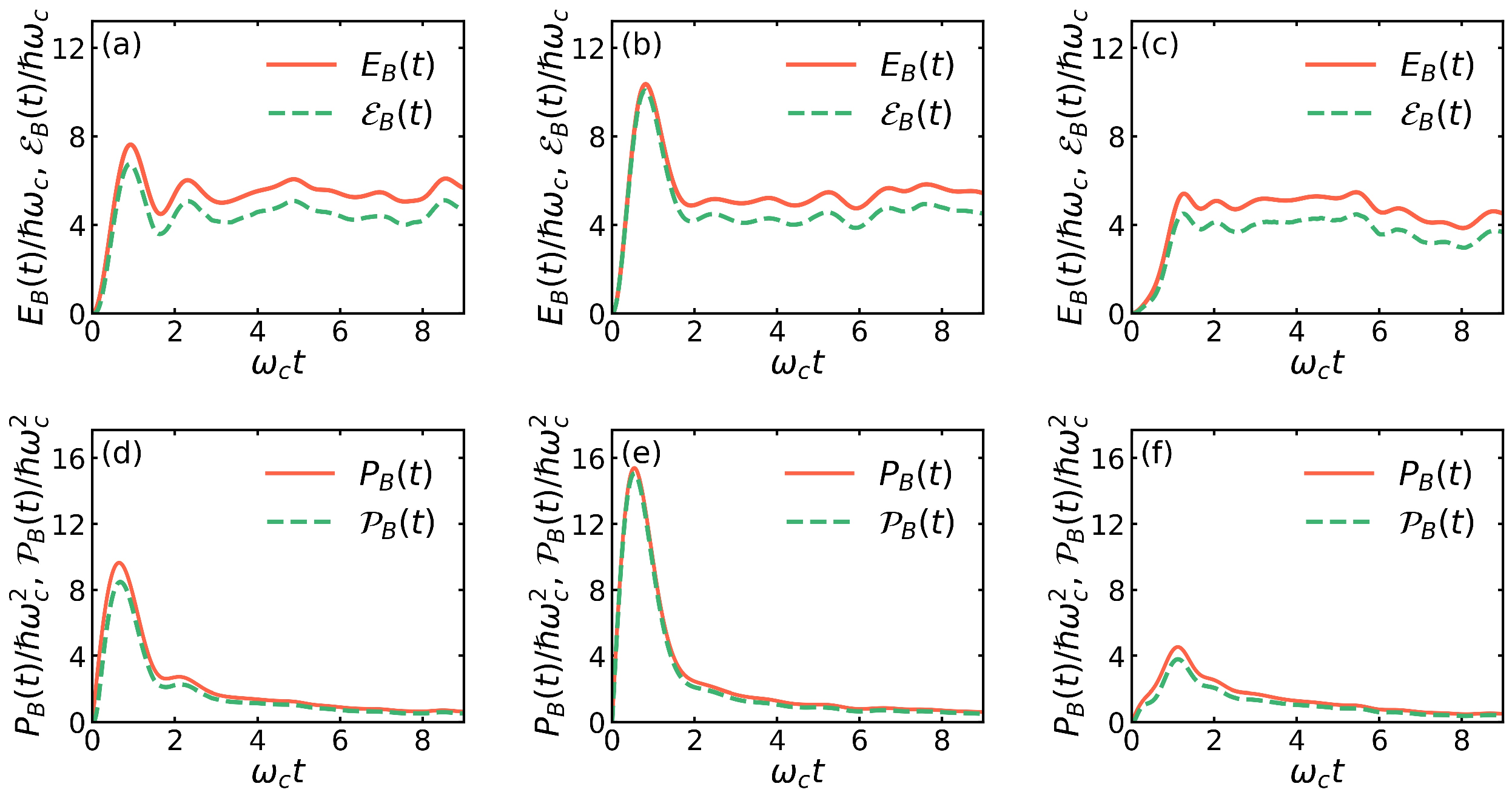}
\caption{\label{fig.2}(a)-(c) The dependence of the stored energy $E_{B}(t)$ and ergotropy $\mathcal{E}_{B}(t)$ (both in units of $\hbar\omega_{c}$) as a function of $\omega_{c}t$ for three selected states of the charger: a Fock state, a coherent state, and a squeezed state, respectively. Panels (d)-(f) depict the variation of the average charging powers, $P_{B}(t)$ and $\mathcal{P}_{B}(t)$ (both in units of $\hbar\omega_{c}^2$) on $\omega_{c}t$ for the same three charger states, respectively. In the following calculations, we use the same settings of $N=6$, $g_{12}= g_{23}=1$, unless otherwise specified.}
\end{figure*}

The ratio between ergotropy and stored energy as another useful quantifier of QB performance, is defined as
\begin{eqnarray}\label{R(t)}
R_{B}(t)=\mathcal{E}_{B}(t)/E_{B}(t).
\end{eqnarray}

Due to the unitary evolution of the QB system during charging, there will be a reciprocal exchange of energy between the charger and the battery. It may not be necessary to continuously monitor the stored energy, ergotropy and charging power of the QB throughout the charging process. Therefore, we choose the maximum stored energy $E_{max}$ (at time $t_{E}$), maximum ergotropy $\mathcal{E}_{max}$ (at time $t_\mathcal{E}$), and maximum charging powers $P_{max}$ (at time $t_{P}$) and $\mathcal{P}_{max}$ (at time $t_{\mathcal{P}}$), to measure QB performance,
\begin{equation}\label{Emax}
E_{max}\equiv \mathop{max}\limits_{t}[E_{B}(t)]=E[(t_{E})],
\end{equation}
\begin{equation}\label{emax}
\mathcal{E}_{max}\equiv \mathop{max}\limits_{t}[\mathcal{E}_{B}(t)]=\mathcal{E}[(t_\mathcal{E})],
\end{equation}
\begin{equation}\label{Pmax}
P_{max}\equiv \mathop{max}\limits_{t}[P_{B}(t)]= P[(t_{P})],
\end{equation}
\begin{equation}\label{Pebmax}
\mathcal{P}_{max}\equiv \mathop{max}\limits_{t}[\mathcal{P}_{B}(t)]= \mathcal{P}[(t_{\mathcal{P}})].
\end{equation}
In addition, the entanglements $\overline{S}_{t_E}$ (at time $t_{E}$) and $\overline{S}_{t_P}$ (at time $t_{P}$) corresponding to the maximum stored energy and maximum charging power, respectively, are considered. These measures are given by
\begin{equation}\label{Ste}
\overline{S}_{t_E}\equiv S(t_{E}),
\end{equation}
\begin{equation}\label{Stp}
\overline{S}_{t_P}\equiv S(t_{P}).
\end{equation}

In what follows, we shall analyze the quantities $E_{max}$, $\mathcal{E}_{max}$, $P_{max}$, and $\mathcal{P}_{max}$, as well as their ratios
\begin{equation}\label{RE}
\mathcal{R}_{e}=\mathcal{E}_{max}/E_{max},
\end{equation}
and
\begin{equation}\label{RP}
\mathcal{R}_{p}=\mathcal{P}_{max}/P_{max}.
\end{equation}

In all subsequent calculations, we choose the energy spectrum of our three-level system as $\omega_{1}=0$, $\omega_{2}=1$ and $\omega_{3}=1.95$. We take $\omega_{c}$ as a dimensionless parameter and let $\omega_{c}=1$. Numerical work has been performed by using the PYTHON toolbox QuTiP2 \cite{JOHANSSON20131234}.
\section{THE CHARGING PROPERTY} \label{section3}
In this section, we focus on the charging process of the QB with three initial states of the charger, i.e., a Fock state, a coherent state, and a squeezed state. We then analyze the roles of the entanglement between the charger and the battery. Furthermore, we discuss the asymptotic freedom of the extractable energy and corresponding charging power in the QB.
\subsection{Charging properties and entanglement}\label{subsection1}
In order to analyze the effect of the three different initial states of the charger on the QB, we illustrate the time evolution of the stored energy, ergotropy, and average charging powers as shown in Fig.~\ref{fig.2}. It demonstrates that when the cavity is in a coherent input state, a QB can achieve higher stored energy, ergotropy, and average charging powers than those achieved with Fock or squeezed input states, different from the results obtained in Ref. \cite{PhysRevLett.122.047702,delmonte2021characterization}. In particular, when the charger is initially in a coherent state, almost all the stored energy of the battery can be extracted to generate valuable work at short times. Therefore, for coherent states, the charging power corresponds to the ergotropy, converges to the average storing power in short term. Furthermore, there is a significant difference between the stored energy and ergotropy over long periods of time for any given initial states of the charger.
\begin{figure}[htbp]
\centering
\includegraphics[width=0.36\textwidth]{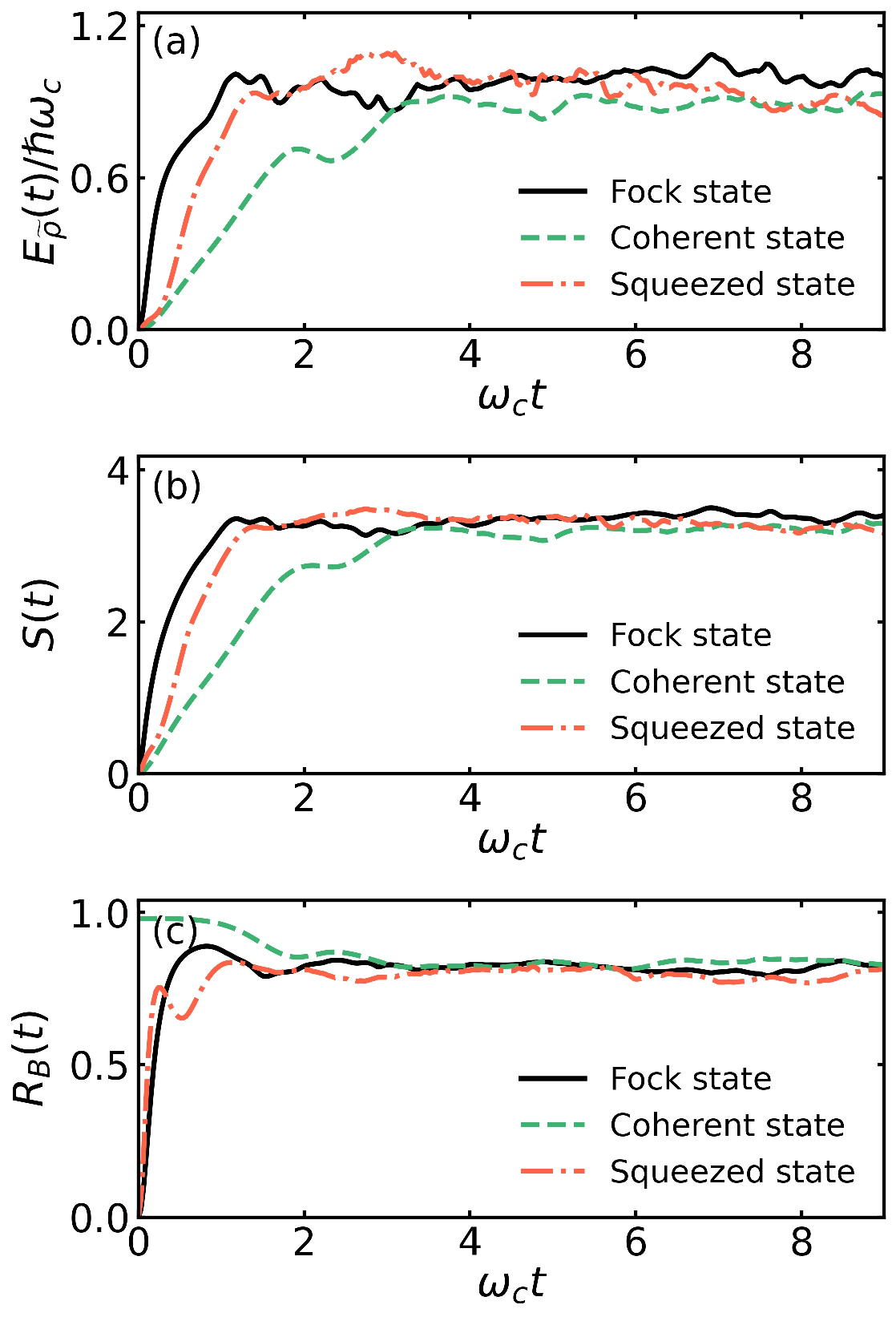}
\caption{\label{fig.3}(a) The locked energy $E_{\widetilde{\rho}}(t)$ (in units of $\hbar\omega_{c}$) as a function of $\omega_{c}t$ for three initial states of the charger. (b) The entanglement $S(t)$ and (c) the ratio $R_{B}(t)$ as a function of $\omega_{c}t$ for the same initial states as in panel (a).}
\end{figure}

We investigate the time evolution of the locked energy $E_{\widetilde{\rho}}(t)$, the entanglement $S(t)$, and the ratio $R_{B}(t)$ for different initial states as shown in Fig.~\ref{fig.3}. The locked energy and the entanglement show an obvious consistent behavior for all three states. Compared to Fock and squeezed states, coherent states take a longer time to generate more entanglement, which is also reflected in the locked energy. This can be attributed to the fact that a more mixed battery state produces more entanglement between the charger and the battery, leading to an increase in the amount of locked energy, which makes it more difficult to extract the battery energy. When entanglement is relatively small, a coherent state allows for the almost complete extraction of stored energy as valuable work. Moreover, a nearly stable ratio can be achieved across the three different initial states of the charger when entanglement is significant. By comparing Fig.~\ref{fig.2} (a)-(c) with Fig.~\ref{fig.3}, it is desirable to minimize the entanglement between the charger and the battery to maximize the capability of work extraction. Such characteristics indicate that Fock and squeezed states induce a complex and entangling dynamics, resulting in a higher degree of entanglement that may lead to a suppression of extractable energy. Conversely, coherent states may be optimal in the sense of energy extraction because they produce a smaller amount of entanglement, which is consistent with the conclusion demonstrated by Ref. \cite{PhysRevLett.122.047702}.
\begin{figure}[htbp]
\centering
\includegraphics[width=0.49\textwidth]{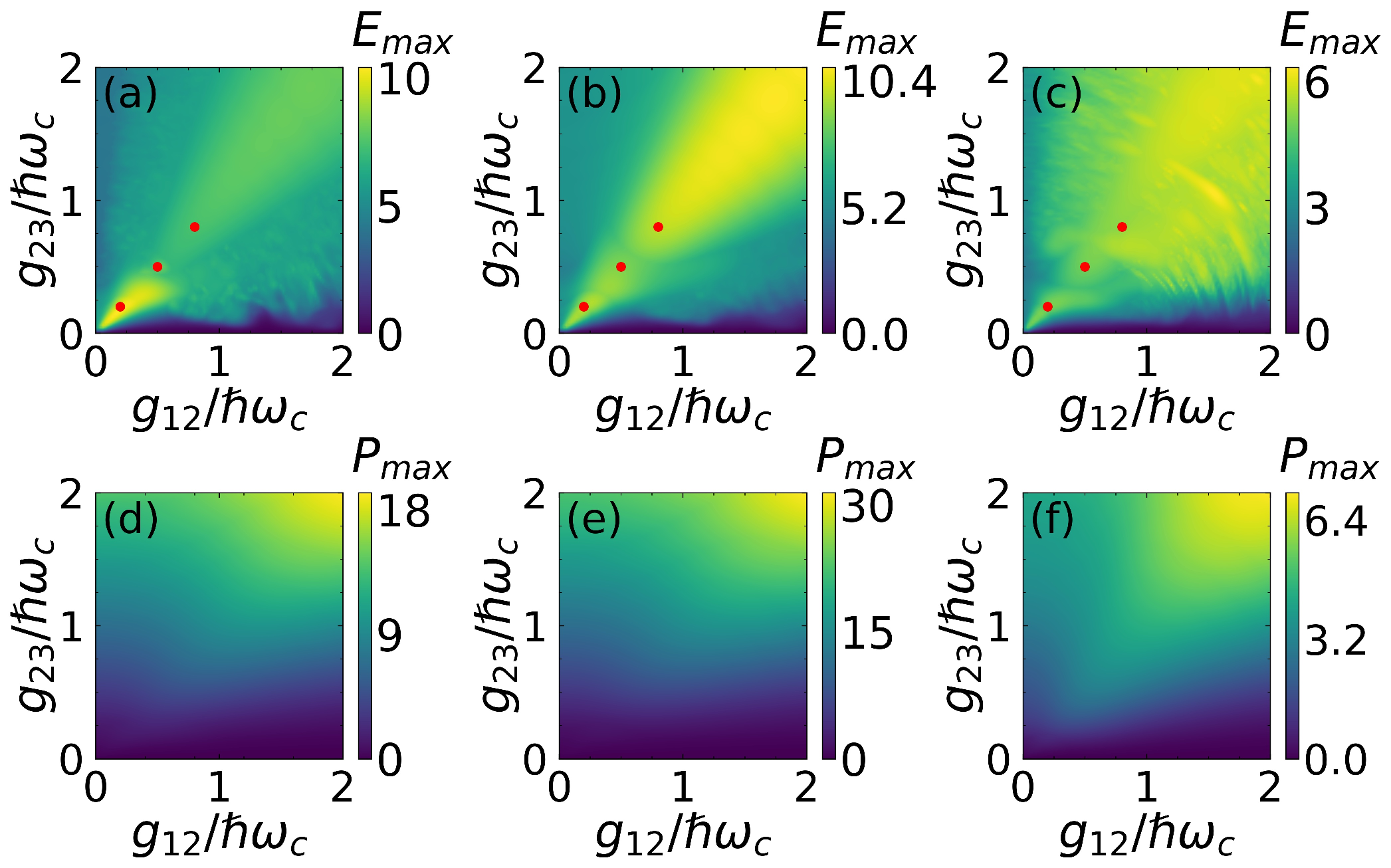}
\caption{\label{fig.4}The contour plots of the QB's maximum stored energy $E_{max}$ (in units of $\hbar\omega_{c}$) [panels (a)-(c)] and maximum charging power $P_{max}$ (in units of $\hbar\omega_{c}^2$) [panels (d)-(f)] with different initial states of charger. The three initial states of charger are as follows: (a) and (d): a Fock state, (b) and (e): a coherent state, and (c) and (f): a squeezed state, respectively. Here, we consider the ranges $g_{12}\in[0,2]$, $g_{23}\in[0,2]$. Both $g_{12}$ and $g_{23}$ are in units of $\hbar\omega_{c}$. The red dots are three randomly selected points near the position where a change of quantum phases occurs. The values of $g_{12}$ and $g_{23}$ for different points are as follows: $(0.2, 0.2)$, $(0.5, 0.5)$, and $(0.8, 0.8)$.}
\end{figure}

The results of the maximum stored energy and the maximum charging power of the QB as a function of the cavity-atom coupling strengths $g_{12}$ and $g_{23}$ are shown in Fig.~\ref{fig.4}. It further shows that the coherent state as the initial state of the charger is the optimal protocol to achieve higher maximum stored energy and maximum charging power. The maximum stored energy of the QB exhibits a distinct approximate ``triangular" region when coupling strengths $g_{12}$ and $g_{23}$ in a closer range under three different initial states. In the ``triangular" region, quantum phase transitions of the QB system lead to a variation in the maximum stored energy across different phases (see Appendix \ref{appendix1} for details) \cite{PhysRevA.86.063822,PhysRevA.87.023813,
PhysRevA.87.023805,PhysRevA.107.033711}. We randomly select three points near the change of quantum phases within the ``triangular" region. For a Fock state, the QB in the normal phases can achieve a greater maximum energy storage compared to the superradiant phases. Conversely, a coherent and squeezed states in the superradiant phases obtain the higher maximum stored energy. The maximum stored energy and maximum charging power of the QB in the squeezed state are less than those of the other two states. In addition, stronger cavity-atom coupling strengths in the ``triangular" region increase the maximum charging power for all three initial states of the charger.
\begin{figure}[htbp]
\centering
\includegraphics[width=0.48\textwidth]{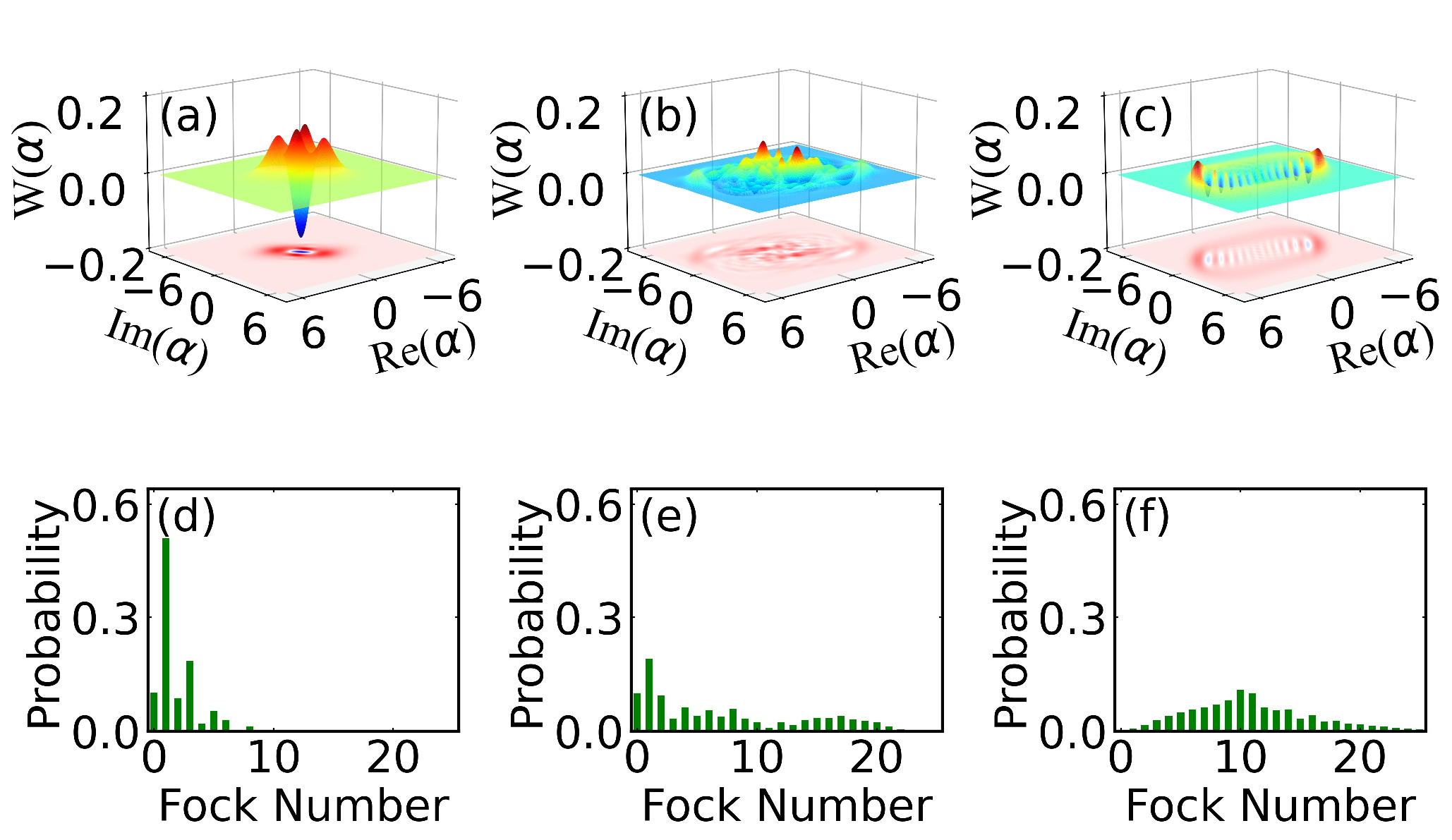}
\caption{\label{fig.5}(a)-(c) Wigner function and (d)-(f) photon distribution in Fock basis corresponding to the maximum stored energy of the charger in Fock states for different cavity-atom coupling strengths. The values of $g_{12}$ and $g_{23}$ are as follows: (a) and (d): $(0.2, 0.2)$, (b) and (e): $(0.5, 0.5)$, (c) and (f): $(0.8, 0.8)$.}
\end{figure}

To further clarify the behavior of the maximum stored energy, we introduce the Wigner function (see Appendix \ref{appendix2} for details) and the photon distribution in Fock basis \cite{PhysRev.40.749,HILLERY1984121,Kim1990,Weinbub2018,PhysRev.131.2766}. Figure~\ref{fig.5} illustrates the cavity Wigner function and photon distribution corresponding to the maximum stored energy in Fock states at three selected points (see Fig.~\ref{fig.4}). At small cavity-atom coupling strengths, the significantly negative value of the Wigner function indicates the presence of nonclassical properties and the probability distribution dominated by odd photons in the range of photon numbers 10, enabling the QB to obtain a higher stored energy. With the coupling further enhanced, the negative value of the Wigner function decreases, and the Wigner function is completely separated into two peaks. The probability becomes more evenly distributed between odd and even number of photons and eventually satisfies a Gaussian profile, which means that the charger can supply more energy to the battery. And then the energy obtained by the QB in the superradiant phases also gradually increases as the coupling strengthens. Similarly, we also calculate the Wigner function and the photon distribution for coherent and squeezed states in Appendix \ref{appendix2}.

\subsection{Asymptotic freedom} \label{subsection2}
We have investigated how the entanglement and cavity-atom coupling strengths affect the charging process of the QB when the number of batteries is fixed for three initial states of the charger. Hereafter, we will analyze the effect of the entanglement on the QB's
performance as the number $N$ of three-level atoms increases.
\begin{figure}[htbp]
\centering
\includegraphics[width=0.48\textwidth]{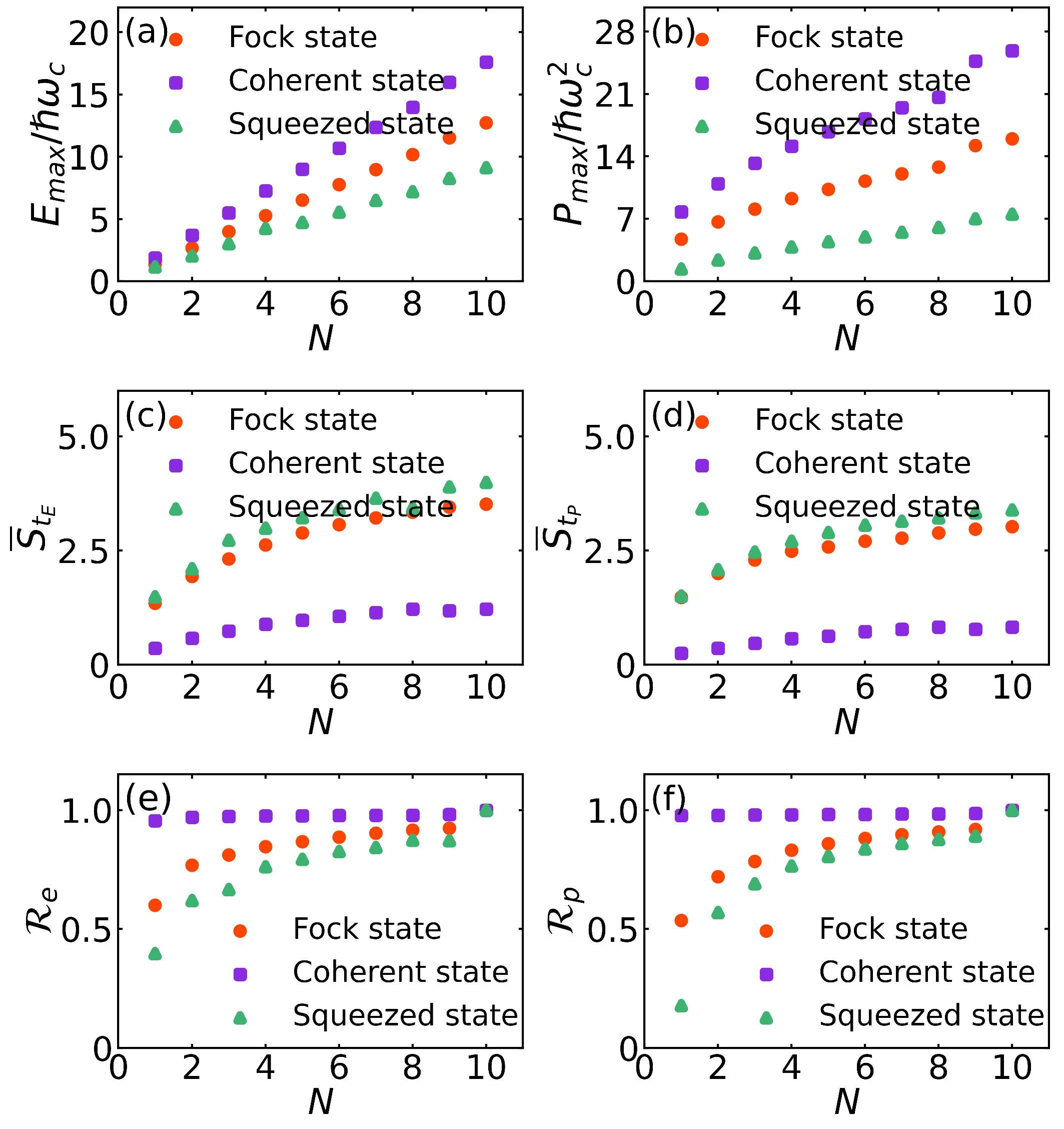}
\caption{\label{fig.6}(a) and (b) The maximum stored energy $E_{max}$ (in units of $\hbar
\omega_{c}$) and the maximum charging power $P_{max}$ (in units of $\hbar\omega_{c}^2$) as a function of $N$ for different initial states of the charger, respectively. (c) and (d) The entanglements $\overline{S}_{t_E}$ and $\overline{S}_{t_P}$ as a function of $N$ for the same states as in panel (a), individually. Panels (e)-(f) depict the variation of the ratios, $\mathcal{R}_{e}$ and $\mathcal{R}_{p}$ on $N$ for different initial states same as in panel (a), respectively. Here, we consider the ranges $N\in[0,10]$.}
\end{figure}

In Fig.~\ref{fig.6}, we display the maximum stored energy $E_{max}$, the maximum charging power $P_{max}$, the entanglements $\overline{S}_{t_E}$ and $\overline{S}_{t_P}$, and the ratios $\mathcal{R}_{e}$ and $\mathcal{R}_{p}$ as a function of $N$ for three initial states of the charger. A coherent state as the initial state for the charger, the QB can obtain a greater maximum stored energy and a higher maximum charging power as $N$ increases, while maintaining a smaller entanglement between the charger and the battery than Fock and squeezed states. We naturally conclude that the entanglement is main limiting factor in improving the QB's performance for three initial states of the charger. Particularly, the entanglement entropy of the subsystem exhibits a notable characteristic: it grows logarithmically with the size of the battery, thereby adhering to an $area$ $law$ \cite{PhysRevA.66.042327,PhysRevLett.90.227902,RevModPhys.82.277}. However, the stored energy scales linearly with the battery's size. Therefore, we expect that the energy locked up by the entanglement will become negligible compared to the stored energy as $N$ approaches infinity. When the number of three-level atoms $N$ is small and the initial state of the charger is either a Fock or a squeezed state, the locked energy occupies a larger proportion of the stored energy in the battery. Consequently, the extractable energy from the battery is much less than the stored energy, and the fraction of charging power corresponding to the extractable energy is also notably low when $N$ is small, see Fig.~\ref{fig.6}(e) and \ref{fig.6}(f). The coherent state produces less entanglement than the Fock and squeezed states, leading to the lower locked energy. Therefore, the coherent state has a higher extractable energy and corresponding charging power even at small $N$. As $N$ increases, the locked energy becomes much smaller relative to the stored energy, resulting in a larger extractable energy in the battery for Fock and squeezed states. Interestingly, regardless of the initial state of the charger, the energy stored in the QB can be completely extracted when $N=10$, and the behavior of the charging power aligns with the energy as well. We also argue that as the size of our battery continue to grow, the extractable energy will converge to the stored energy, and the charging power associated with the ergotropy will also approach the maximum of the average storing power.
\section{CONCLUSIONS}\label{section4}
We have introduced the concept of the three-level Dicke QB, consisting of $N$ identical and independent $\Xi$-type three-level atoms coupling to a single-mode cavity. We have analyzed the influence of three quantum optical states (Fock, coherent, and squeezed states) of the charger on the performance of QBs, including the stored energy, ergotropy, average charging power. Our results show that the coherent state can significantly enhance the charging performance compared to the Fock and squeezed states. The locked energy and the entanglement between the charger and the battery exhibit consistent behavior, indicating that entanglement is the main limiting factor in the capability of work extraction. When the level of entanglement is relatively low, a coherent state can almost completely extract the stored energy at short times. The effect of the cavity-atom coupling strengths on the maximum stored energy and the maximum charging power depends on the initial states of the charger. By analyzing the quantum phase transitions, Wigner function, and photon distribution, we have explained the behavior of the maximum stored energy. We have also investigated how the number $N$ influences the maximum stored energy, maximum charging power, entanglements and the ratios. The coherent state leads to a greater maximum stored energy and maximum charging power, while also maintaining a lower entanglement than the Fock and squeezed states. We have found that for small $N$, when the charger is in either the Fock or the squeezed state, the ergotropy can be much less than the energy stored in the battery. The fraction of charging power corresponding to the ergotropy is also considerably low and the coherent state is the optimal initial state. For large values of $N$, almost all of the stored energy in the battery can be extracted. In particular, no matter what the initial state is, all of the stored energy becomes extractable for $N=10$, and the charging power follows the same trend. Our study demonstrated that the extractable energy and corresponding charging power in the QB is asymptotically free as $N$ $\rightarrow$ $\infty$.
\section*{Acknowledgments}
The work is supported by the National Natural Science Foundation of China (Grant No. 12075193).

\appendix
\section{Ground-state energy} \label{appendix1}
\begin{figure}[htbp]
\centering
\includegraphics[width=0.30\textwidth]{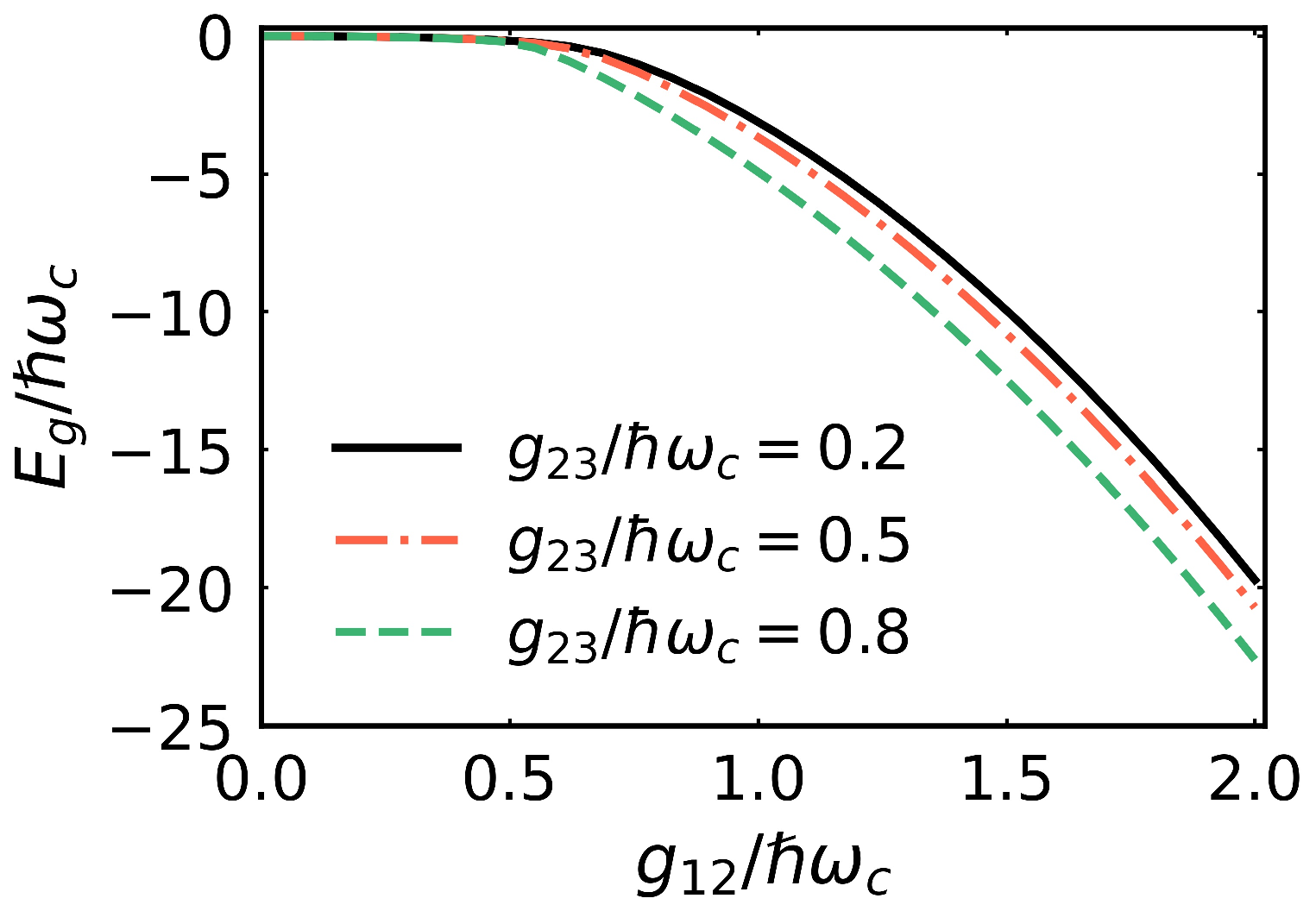}
\caption{\label{fig.7} The dependence of the ground-state energy $E_{g}$ (in units of $\hbar\omega_{c}$) on cavity-atom coupling strength $g_{12}$ (in units of $\hbar\omega_{c}$) for different values of $g_{23}$ (in units of $\hbar\omega_{c}$).}
\end{figure}
Figure~\ref{fig.7} displays the ground-state energy $E_{g}$ as a function of the cavity-atom coupling strength $g_{12}$ for different values of $g_{23}$. The presence of the quantum phase transitions from normal phases ($E_{g}=0$) to superradiant phases ($E_{g}<0$) can be clearly seen in the calculation of the ground-state energy. When the values of the coupling strengths $g_{12}$ and $g_{23}$ are $(0.5, 0.5)$, the quantum phase changes significantly.
\section{Wigner function and photon distribution in coherent and squeezed states} \label{appendix2}
A convenient way to visualize a quantum state is through the Wigner function using the phase-space formalism \cite{PhysRev.40.749}, which can be defined as the Fourier transform of the symmetrically ordered characteristic function $\chi(\eta)$
\begin{eqnarray}\label{wigner function}
\begin{split}
W(\alpha)=\frac{1}{\pi^{2}} \int \exp \left(\eta^{*} \alpha-\eta \alpha^{*}\right) \chi(\eta) \mathrm{d}^{2} \eta,
\end{split}
\end{eqnarray}
where $\alpha$ is an arbitrary point in phase-space. The symmetrically ordered characteristic function $\chi(\eta)$ is given by
\begin{eqnarray}\label{characteristic function}
\begin{split}
\chi(\eta)=\operatorname{Tr}\left\{\rho_{A} \mathrm{e}^{\eta \hat{a}^{\dagger}-\eta^{*} \hat{a}}\right\},
\end{split}
\end{eqnarray}
$\rho_{A}$ is the reduced density matrix of the charger in our calculations. The Wigner function also can be defined in terms of the generalized conjugate position $x$ and momentum $p$ as $\bar{W}(x, p)=\frac{1}{\pi \hbar} \int_{-\infty}^{\infty}\langle x+z|\rho_{A}| x-z\rangle e^{-2 i p z / \hbar} d z$, and $W(\alpha)$ can be derived from $\bar{W}(x, p)$.
\begin{figure}[htbp]
\centering
\includegraphics[width=0.48\textwidth]{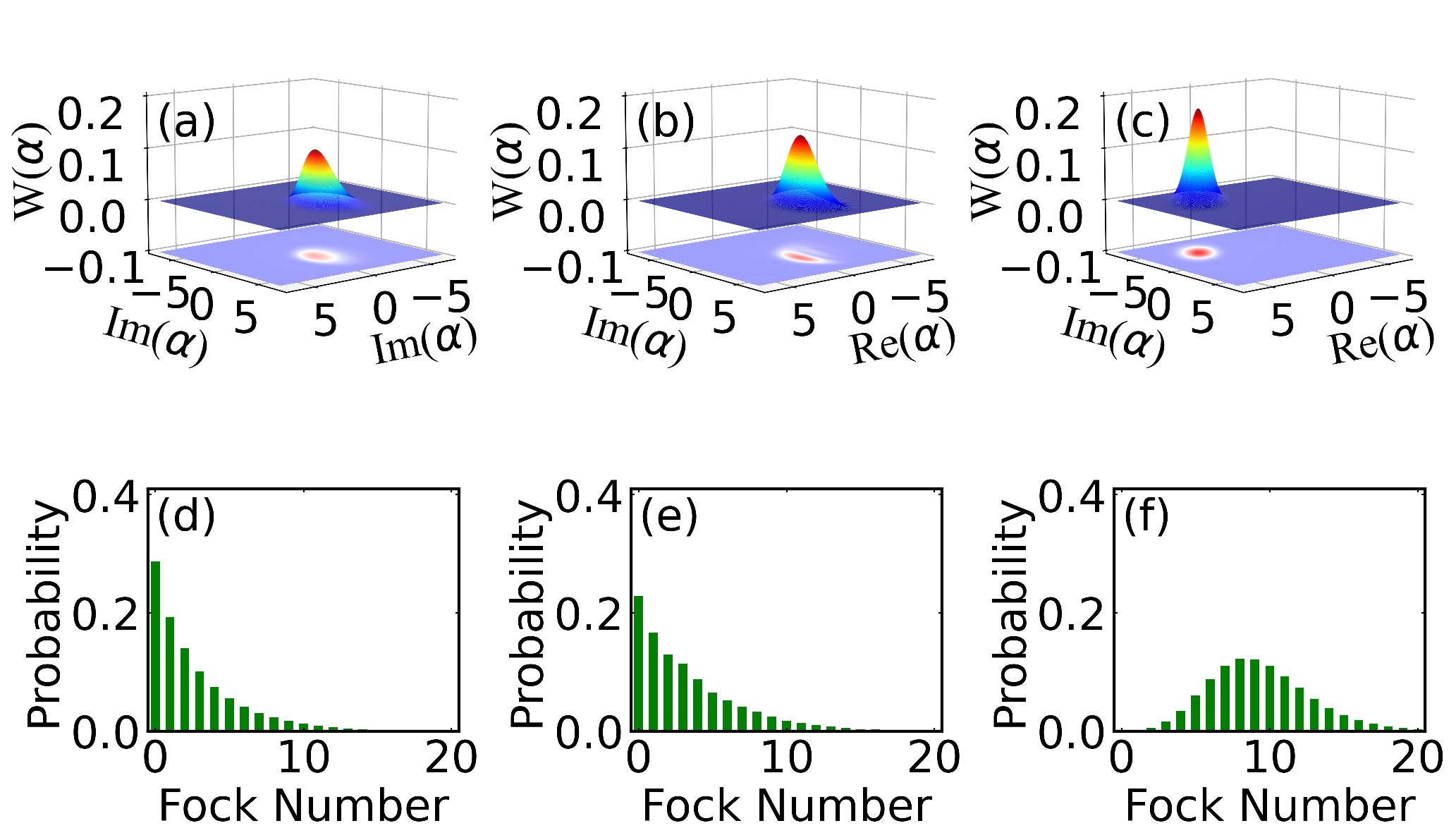}
\caption{\label{fig.8}(a)-(c) Wigner function and (d)-(f) photon distribution in Fock basis corresponding to the maximum stored energy of the charger in coherent states for different cavity-atom coupling strengths. The values of $g_{12}$ and $g_{23}$ are as follows: (a) and (d): $(0.2, 0.2)$, (b) and (e): $(0.5, 0.5)$, (c) and (f): $(0.8, 0.8)$.}
\end{figure}

\begin{figure}[htbp]
\centering
\includegraphics[width=0.48\textwidth]{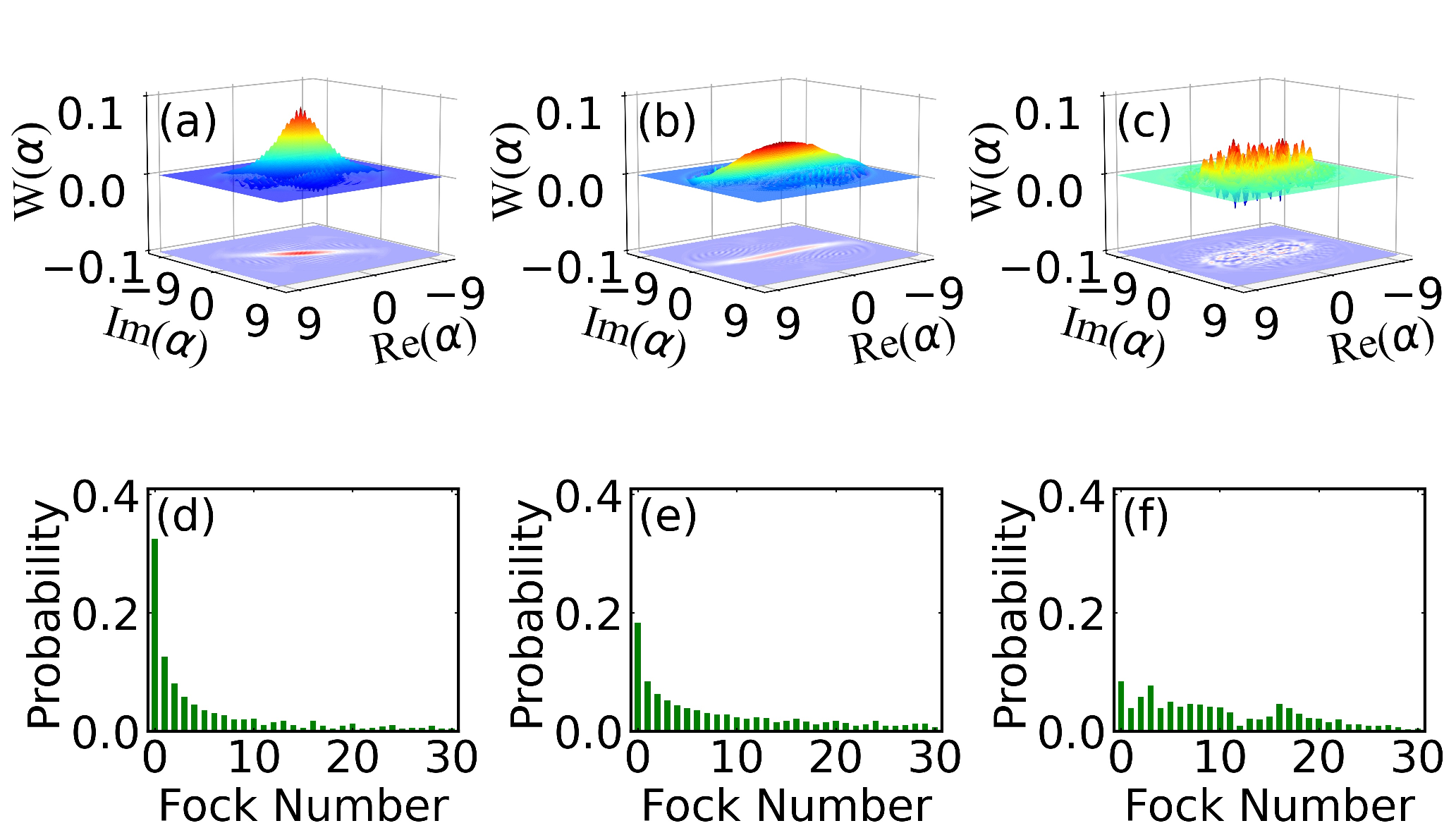}
\caption{\label{fig.9}(a)-(c) Wigner function and (d)-(f) photon distribution in Fock basis corresponding to the maximum stored energy of the charger in squeezed states for different cavity-atom coupling strengths. The values of $g_{12}$ and $g_{23}$ are as follows: (a) and (d): $(0.2, 0.2)$, (b) and (e): $(0.5, 0.5)$, (c) and (f): $(0.8, 0.8)$.}
\end{figure}

We calculate the Wigner function and photon distribution corresponding to the maximum stored energy for the charger in coherent and squeezed states, are shown in Figs.~\ref{fig.8} and \ref{fig.9}, respectively. As the coupling strengths increase, the Wigner function of the coherent state shows ``ripples" and
the photon distribution extends towards larger photon numbers. At stronger coupling, the Wigner function transforms into an oval shape, and the photon probability distribution follows a Gaussian profile. Therefore, for a coherent state, the QB in the superradiant phases achieves a higher maximum stored energy. For the squeezed state, its Wigner function displays prominent oscillatory behavior and distinct ``ripples" as well. Stronger coupling strengths improve the negative value of the squeezed state's Wigner function, leading to an enhancement of the maximum stored energy. The photon distribution in the squeezed state oscillates under strong coupling, resulting in minimal stored energy in comparison to the Fock and coherent states.
\bibliography{reference}
\end{document}